\documentclass[pdflatex,sn-mathphys-num]{sn-jnl}% Math and Physical Sciences Numbered Reference Style 
\usepackage{graphicx}%
\usepackage{multirow}%
\usepackage{amsmath,amssymb,amsfonts}%
\usepackage{amsthm}%
\usepackage{mathrsfs}%
\usepackage[title]{appendix}%
\usepackage{xcolor}%
\usepackage{textcomp}%
\usepackage{manyfoot}%
\usepackage{booktabs}%
\usepackage{algorithm}%
\usepackage{algorithmicx}%
\usepackage{algpseudocode}%
\usepackage{listings}%
\begin{document}

\title[Dispersive vacuum as a decoherence amplifier of an Unruh-DeWitt detector]{Dispersive vacuum as a decoherence amplifier of an Unruh-DeWitt detector}


\author*{\fnm{Pedro H. M.} \sur{Barros}*}\email{phmbarros@ufpi.edu.br}

%\author{\fnm{Helder A. S.} \sur{Costa}}\email{hascosta@ufpi.edu.br}

\author{\fnm{Helder A. S.} \sur{Costa}}\email{hascosta@ufpi.edu.br}

\affil{\orgdiv{Departamento de Física}, \orgname{Universidade Federal do Piauí}, \orgaddress{\street{Ininga}, \city{Teresina}, \postcode{64049-550}, \state{Piauí}, \country{Brasil}}}

%\affil[2]{\orgdiv{Department}, \orgname{Organization}, \orgaddress{\street{Street}, \city{City}, \postcode{10587}, \state{State}, \country{Country}}}

%\affil[3]{\orgdiv{Department}, \orgname{Organization}, \orgaddress{\street{Street}, \city{City}, \postcode{610101}, \state{State}, \country{Country}}}


\abstract{Recently, interest has been growing in studies on discrete or ``pixelated'' space-time that, through modifications in the dispersion relation, can treat the vacuum as a dispersive medium. Discrete spacetime considers that spacetime has a cellular structure on the order of the Planck length, and if this is true we should certainly have observable effects. In this paper, we investigated the effects caused by the dispersive vacuum on the decoherence process of an Unruh-DeWitt detector, our setup consists of a uniformly accelerated detector, initially in a qubit state, which interacts with a massless scalar field during a time interval finite. We use dispersion relations drawn from doubly special relativity and Hořava-Lifshitz gravity, with these modifications the vacuum becomes dispersive and has a corresponding refractive index. We calculate the probability transition rates, the probability of finding the detector in the ground state, and the quantum coherence variation. Our results indicate that the decoherence process occurs more quickly in cases with changes in the dispersion relation in the regime of high accelerations and interaction time. Additionally, the decoherence increases as the vacuum becomes more dispersive due to the increase in the order of modification in the dispersion relation, and this happens because the dispersive vacuum amplifies the effects of quantum fluctuations that are captured by the detector when interacting with the field.}




\maketitle

\section{\label{sec:level1} Introduction}

Quantum field theory in curved spacetime presents fundamental and fascinating challenges at the frontier of scientific knowledge in theoretical physics. In this context, a prominent phenomenon is the so-called Unruh Effect or Fulling-Davies-Unruh Effect \cite{unruh1976, fulling1973, Davies_1975}. According to this effect, an observer accelerating through the quantum vacuum will perceive a thermal bath of particles at a temperature proportional to their acceleration \cite{matsas2008}. Although this effect may seem non-intuitive, it makes perfect sense if the idea of a vacuum is correctly interpreted, and therefore, the vacuum plays a very important role here.
The Unruh-DeWitt detector \cite{unruh1976, DeWitt1980} comes into play as a conceptual tool for exploring and measuring these particles \textquotedblleft created" during acceleration, and so, it can be said that \textquoteleft particles are what the particle detector detects' \cite{davies1984}. Something important about the Unruh-DeWitt detector is that in the particle detection process, it reveals the most important aspects of how atoms interact with the electromagnetic field \cite{Martinez2013wavepacket, Martinez2018, Martinez2021}. The initial study of the Unruh effect used an in-out method, assuming that the initial state of the system was defined in the infinite past and the final state evaluated in the infinite future. Unlike this, some studies, such as \cite{svaiter1992, Higuchi1993matsas, sriramkumar1996finite, Shevchenko2017}, has delved into how uniformly accelerated detectors respond when interacting with a scalar field for a finite time.

Another way to detect this creation of particles indirectly is through quantum coherence, which refers to the ability of a quantum system to maintain superposed states. Theoretically, the relationship between quantum coherence and the Unruh Effect is studied through models that incorporate the principles of quantum field theory and quantum mechanics, these models attempt to predict how coherent dynamics evolve in relativistic environments, an example would be a single-qubit accelerated. When subjected to the Unruh Effect, the temporal evolution of these superposed states can be altered and the presence of created particles can introduce factors that accelerate the degradation of coherence, as reported in some studies \cite{Huang2018, Huang2019, Nesterov2020, Zhang2022, huang2022, Pedro2024}. An interesting curiosity also trained is that coherence can be maintained, avoidance, mitigation, and revived, as can be seen in \cite{Unruh1995, Schlosshauer2019, feng2021}. Direct detection of the Unruh Effect and its effects on quantum coherence demands precise experimental conditions, often exceeding current technological capacity.

Historically, several studies \cite{hossenfelder2013minimal, trugenberger2016random, trugenberger2017combinatorial, tee2020dynamics, tee2022fundamental} have indicated that spacetime emerges at the macroscopic limit from some type of discrete or pixelated substructure, and therefore, the vacuum can be considered as a dispersive medium. According to quantum gravity, this pixel is on the order of the Planck length scale ($10^{-35}$ m). This type of length scale violates Lorentz invariance, but through the construction known as doubly special relativity (DSR) \cite{AMELINOCAMELIA2001, camelia2002, Ambjorn2002}, Minkowski space can be maintained if the momentum space has non-zero curvature. We can modify the dispersion relation by adding higher order terms, this type of modification is found in the Hořava-Lifshitz \cite{HoravaLifshitz2009} theories of gravity (HLG), which uses an anisotropic scale of space and time by introducing a scale parameter and a critical exponent. Therefore, these studies have shown that it is possible to obtain dispersive vacuum by modifying the dispersion relation, this modification provides a refractive index for the vacuum, which is now dispersive, on the other hand, this produces light waves of different frequencies propagating with different speeds.

With the aim of investigating the observable effects of discrete spacetime, recently, a study \cite{PaulDavies2023} investigated how an accelerated detector interacting with a massless scalar field is affected when the dispersion relation is modified. Precisely, they showed that modifications in the dispersion relation present in DSR and HLG can influence the response of the Unruh-DeWitt detector. Based on these results, in this paper we probe the robustness of the dispersive vacuum effect on the degradation of the quantum coherence of a single-qubit accelerated during a finite interaction time, to understand the observable consequences of the dispersive vacuum (caused by changes in the dispersion relation) in two-level quantum systems.

This article is structured as follows: Initially, in Sec. \ref{sec:level2} we present the physical model that describes the interaction of the massless scalar field with the accelerated detector providing the transition probability rate and we study this same model for cases with dispersive vacuum as well as its numerical results. Then, Sec. \ref{sec:level3} presents the quantum setup to estimate the decoherence of a single-qubit induced by the Unruh effect in a dispersive vacuum. Through this scheme, we calculate the probability of finding the two-level system in the ground state and the $l^1$ norm quantum coherence variation as well as their respective numerical results as a function of dimensionless parameters. Finally, in the final section, we present the conclusions expressed by the discussion of our results. Here we use the system of natural units in which $\hbar = c = k_B = 1$.

\section{\label{sec:level2} Accelerated detectors with dispersive vacuum}

In the interaction picture, the effective coupling between the detector with internal states $|g\rangle$ and $|e\rangle$ and a massless scalar field $\phi$, in the Minkowski vacuum $|0_{\mathcal{M}}\rangle$, is described by the interaction Hamiltonian $\hat{\mathcal{H}}_{\mathrm{int}}$ in the detector frame
\begin{equation}\label{Hint}
\mathcal{H}_{\mathrm{int}} =  g\Theta(\tau)\mu(\tau)\phi[x(\tau)],
\end{equation}
where we assume that $\tau$ is the proper time of the detector that follows a trajectory $x(\tau)$, $\mu(\tau)$ is the monopole moment of detector, $g$ is small and denotes the effective coupling constant between the detector and the field, and $\Theta(\tau)$ is a gradual window function used to describe finite time interaction, it has the following properties: $\Theta(\tau) \approx 1$ for $|\tau| \ll T$ and  $\Theta(\tau) \approx 0$ for $|\tau| \gg T$, where $T$ is the interaction time. The probability of excitation and emission from the detector (to the first order in perturbation theory) is given by
\begin{align} \label{P1}
  \mathcal{P}^{-} = g^2|\langle g|\mu(0)|e\rangle|^2 \mathcal{F}^{-}, \\
  \mathcal{P}^{+} = g^2|\langle e|\mu(0)|g\rangle|^2 \mathcal{F}^{+}, \label{P2}
 \end{align}
where the response function $\mathcal{F}^{\pm}$ is given by
\begin{eqnarray} 
\mathcal{F}^{\pm} = \int_{-\infty}^{\infty} d\tau  \int_{-\infty}^{\infty} d\tau' \Theta(\tau)\Theta(\tau') e^{\pm i\Omega(\tau - \tau')} D^{+}[x(\tau), x(\tau')], \label{intF}
\end{eqnarray}
where $\Omega$ is the transition angular frequency with: $\Omega > 0$ and $\Omega < 0$ corresponding to excitation and de-excitation, respectively, and $D^{+}(x(\tau), x(\tau')) = \langle 0_{\mathcal{M}}|\phi[x(\tau)]\phi[x(\tau')]|0_{\mathcal{M}}\rangle$ is the Wightman function \cite{birrell1984quantum}. According to \cite{sriramkumar1996finite}, we have that transition probability depends on the interaction time as given by $\mathcal{F}^\pm \approx \mathcal{F}^\pm (\infty) - \Theta''(0)\frac{\partial^2\mathcal{F}^\pm (\infty)}{\partial\Omega^2}$ where $\mathcal{F}^\pm (\infty)$ corresponds to the infinite-time detector and the respective transition probability per unit time is expressed as $\mathcal{R}^\pm \approx \mathcal{R}^\pm (\infty) - \Theta''(0)\frac{\partial^2\mathcal{R}^\pm (\infty)}{\partial\Omega^2}$, where $\mathcal{R}(\infty) = \int_{-\infty}^{\infty} d\Delta\tau e^{-i\Omega\Delta\tau}G^{+}(\Delta\tau)$ with $\Delta\tau = \tau - \tau'$. We employ the Gaussian window function $\Theta(\tau) = \exp\left(-\frac{\tau^2}{2T^2}\right)$, which provides a gradual and smooth activation and deactivation of the detector. The finite-time adjustments to the transition probability rate can be represented as:
\begin{eqnarray} \label{R}
 \mathcal{R}^\pm \approx \mathcal{R}^\pm (\infty) + \frac{1}{2T^2}\frac{\partial^2\mathcal{R}^\pm (\infty)}{\partial\Omega^2} + \mathcal{O}\left(\frac{1}{T^4}\right).
\end{eqnarray}
where the de-excitation probability rate and the excitation probability rate are, respectively, $\mathcal{R}^{+}$ and $\mathcal{R}^{-}$. For a detector undergoing uniform acceleration with proper acceleration $a$, the intrinsic coordinate system within the detector's frame is connected to the Minkowski coordinates through the transformations \cite{rindler1966kruskal}
\begin{eqnarray}\label{TransfRindler}
    t(\tau) = \frac{1}{a}\sinh\left(a\tau\right), \quad
    x(\tau) = \frac{1}{a}\cosh\left(a\tau\right), \quad
    y(\tau) = z(\tau) = 0.
\end{eqnarray}
and from these transformations the Wightman function can be expressed as $D^{+}(\Delta\tau) = \frac{-1}{4\pi^2}\sum_{k=-\infty}^{\infty}(\Delta\tau - 2i\epsilon + 2\pi ik/a)^{-2}$, and the transition probability rate becomes
\begin{eqnarray} \label{A3}
 \overline{\mathcal{R}}^-_{0} &\approx& \frac{1}{2\pi}\frac{1}{e^{2\pi/\overline{a}} - 1}
 \Bigg\lbrace 1 + \frac{2\pi}{\overline{a}\sigma^2} \frac{e^{2\pi/\overline{a}}}{e^{2\pi/\overline{a}} - 1} \left[1 - e^{2\pi/\overline{a}} + \frac{\pi}{\overline{a}}\left(e^{2\pi/\overline{a}} + 1\right) \right] \Bigg\rbrace,
\end{eqnarray}
\begin{eqnarray} \label{A3b}
 \overline{\mathcal{R}}^+_{0} &\approx& \frac{1}{2\pi}\frac{e^{2\pi/\overline{a}}}{e^{2\pi/\overline{a}} - 1}
 \Bigg\lbrace 1 + \frac{2\pi}{\overline{a}\sigma^2} \frac{1}{e^{2\pi/\overline{a}} - 1} \left[1 - e^{2\pi/\overline{a}} + \frac{\pi}{\overline{a}}\left(e^{2\pi/\overline{a}} + 1\right) \right] \Bigg\rbrace,
\end{eqnarray}
where we have the following dimensionless parameters: $\overline{\mathcal{R}}^\pm = \mathcal{R}^\pm/\Omega$, $\overline{a} = a/\Omega$, and $\sigma = \Omega T$. The zero index in (\ref{A3}) and (\ref{A3b}) means that these equations have no modifications to the dispersion relation. This represents an approximate expression for the transition probability rate when the detector is progressively coupled to a scalar field within a finite time frame $T$. We can see that through expressions (\ref{A3}) and (\ref{A3b}) in the limit $\sigma \to \infty$ we recover the result of the thermal regime characterized by the Bose distribution, with $T_{\mathrm{U}} = a/2\pi$ being the Unruh temperature. This result indicates that the detector of an observer in constant acceleration will perceive thermal radiation with temperature $T_{\mathrm{U}}$.

Although the detector does not observe exactly thermal radiation, the shape of the response function (\ref{A3}) and (\ref{A3b}) is a modification of the thermal radiation observed in the Unruh effect. At a finite interaction time, the detector does not have enough time to reach complete thermal equilibrium with the quantum field, resulting in a slight modification of the ideal thermal radiation \cite{svaiter1992, Higuchi1993matsas, sriramkumar1996finite}. We can relate it to another similar phenomenon, such as obtaining Bose-Einstein condensate through the evaporative cooling method. In this procedure, if the cooling is interrupted prematurely, the system does not reach complete thermal equilibrium \cite{pethick2008bose}.

The standard energy-momentum dispersion relations can be modified by adding powers of momentum $p$, i.e.
\begin{equation}\label{EpGeneral}
    E^2 = p^2 + m^2 + \frac{\kappa_1}{M_p} p^3 + \frac{\kappa_2}{M^2_p} p^4 + \cdots,
\end{equation}
with $M_p$ being the Planck mass and in this condition the dimensionless kappa coefficients are arbitrary. Eq. (\ref{EpGeneral}) alone puts the theory in violation of Lorentz invariance, which can be remedied by changing the momentum measurement \cite{amelino2014planck, mattingly2005modern}.
Indeed, there should be observable effects if the vacuum is dispersive. In (\ref{EpGeneral}) the term $p^3$ produces a correction to the black hole entropy that is usually discarded by setting $\kappa_1$ to zero \cite{AmelinoCamelia2004LQG}. Furthermore, odd-order terms in $p$ are not invariant under parity inversion, these terms are naturally excluded in many physical theories that preserve this parity symmetry. In many dispersion modification theories, the lowest order terms beyond $p^2$ are considered the first significant corrections \cite{colladay1997, Potting_2013}. Considering the conditions for parity invariance, $p^4$ is the first correction term, followed by $p^6$. Higher order terms, such as $p^8$, tend to be suppressed by very small scale factors and have even smaller effects on the observable energies. This justifies the choice to ignore higher order terms, since they contribute negligibly at the energy scales of interest \cite{oller2019brief}.

We now consider the vacuum as a dispersive medium that appears through specific modifications. We use the modified dispersion relation in the presence of additional high-order terms ($\gamma^2 p^4 - \delta^2 p^6$), this type of modification appears in HLG \cite{HoravaLifshitz2009}. Thus, the modified dispersion relation is expressed as
\begin{equation}\label{Ep6}
    E^2 = p^2 + m^2 + \gamma^2 p^4 - \delta^2 p^6,
\end{equation}
where $\gamma^2 = \kappa\eta^2$, $\delta^2 = \kappa\eta^4$, $\kappa$ controls the sign of the correction and is related to unconventional effects that may arise in certain physical contexts, and $\eta$ is a characteristic length scale of the discretized spacetime and is associated with specific properties of the system in consideration. This modification in (\ref{Ep6}) is often used in physical theories that explore extreme regimes, such as in high-energy physics. Note that, when applying the limit $\delta \to 0$ we obtain the dispersion relation modified only by the term ($\gamma^2 p^4$), such modification is the same provided by the DSR. The equation of motion of (\ref{Ep6}) is given by
\begin{equation}
    \frac{\partial^2 \phi}{\partial t^2} - \frac{\partial^2 \phi}{\partial x^2} + \gamma^2\frac{\partial^4 \phi}{\partial x^4} - \delta^2\frac{\partial^6 \phi}{\partial x^6}= 0,
\end{equation}
where the units $E = \omega$ and $p = k$, and we have that the Wightman function is given by
\begin{eqnarray}\label{D+p6}
    D^{+}_{p^6}(\Delta\tau) &=& \frac{-1}{4\pi^2} \sum^{\infty}_{k=-\infty} \frac{\left( 1 + \frac{a^2\gamma^2}{6} - \frac{3a^4\delta'^2}{5}\right)}{(\Delta\tau - 2i\epsilon - 2\pi ik/a)^2} + \frac{1}{4\pi^2} \sum^{\infty}_{k=-\infty} \frac{\left(\gamma^2 + 3a^2\delta'^2\right)}{(\Delta\tau - 2i\epsilon - 2\pi ik/a)^4} +\nonumber\\
    &-& \frac{1}{4\pi^2} \sum^{\infty}_{k=-\infty} \frac{12\delta'^2}{(\Delta\tau - 2i\epsilon - 2\pi ik/a)^6}
\end{eqnarray}
where $\delta'^2 = (1 + \frac{3}{4}\kappa)\delta^2$. The transition probability rate $\mathcal{R}^\pm_{p^6}(\infty)$ is given by
\begin{equation}\label{Rinfty1p6}
    \mathcal{R}^\pm_{p^6}(\infty) = \int_{-\infty}^{\infty} d(\Delta\tau) e^{-i\Omega\Delta\tau}D^{+}_{p^6}(\Delta\tau),
\end{equation}
and that results in
\begin{eqnarray}\label{Rinfty2p6}
    \overline{\mathcal{R}}^-_{p^6}(\infty) &=& \frac{1}{2\pi} \frac{1}{e^{2\pi/\overline{a}}-1} \Bigg\{ 1 + \left( 1 + \overline{a}^2 \right) \frac{\overline{\gamma}^2}{6} + \left[ \frac{1}{5} + \overline{a}^2 - \frac{6\overline{a}^4}{5}\right]\frac{\Omega^4\delta'^2}{2} \Bigg\},
\end{eqnarray}
\begin{eqnarray}\label{Rinfty2p6b}
    \overline{\mathcal{R}}^+_{p^6}(\infty) &=& \frac{1}{2\pi} \frac{e^{2\pi/\overline{a}}}{e^{2\pi/\overline{a}}-1} \Bigg\{ 1 + \left( 1 + \overline{a}^2 \right) \frac{\overline{\gamma}^2}{6} + \left[ \frac{1}{5} + \overline{a}^2 - \frac{6\overline{a}^4}{5}\right]\frac{\Omega^4\delta'^2}{2} \Bigg\},
\end{eqnarray}
with we have the dimensionless parameter $\overline{\gamma} = \Omega\gamma$. When $\delta \to 0$, we are left with the DSR corrections, and therefore we have
\begin{eqnarray}\label{Rinfty2}
    \overline{\mathcal{R}}^{-}_{p^4}(\infty) = \frac{1}{2\pi} \frac{1}{e^{2\pi/\overline{a}}-1} \Bigg[ 1 + \left(1 + \overline{a}^2 \right)\frac{\overline{\gamma}^2}{6} \Bigg],
\end{eqnarray}
\begin{eqnarray}\label{Rinfty2b}
    \overline{\mathcal{R}}^{+}_{p^4}(\infty) = \frac{1}{2\pi} \frac{e^{2\pi/\overline{a}}}{e^{2\pi/\overline{a}}-1} \Bigg[ 1 + \left(1 + \overline{a}^2 \right)\frac{\overline{\gamma}^2}{6} \Bigg].
\end{eqnarray}

For HLG corrections, according to (\ref{R}), calculating the derivatives, rearranging the terms, and replacing $\delta'^2 = (1+3\kappa/4)\delta^2$ and after some simplifications, we obtain
%\begin{eqnarray}\label{d2Rp6}
    %\frac{\partial^2 \mathcal{R}^\pm_{p^6}(\infty)}{\partial\Omega^2} &=& \frac{1}{2\pi} \Bigg( \frac{\Omega}{e^{\frac{2 \pi  c \Omega }{a}}-1} \left[ \kappa\eta^2 + \left( \frac{3a^2}{c^2} + 2\Omega^2\right)\frac{3\kappa^2\eta^4}{4} \right] + \nonumber\\
    %&+& \frac{4\pi c e^{\frac{2\pi\Omega c}{a}}}{a \left(e^{\frac{2\pi\Omega c}{a}}-1\right)^2} \Bigg\{ 1 + \left( \frac{a^2}{3c^2} + \Omega^2\right) \frac{\kappa\eta^2}{2} + \left[ \Omega^2\left( \frac{3a^2}{c^2} + \Omega^2\right) - \frac{6a^4}{5c^4}\right]\frac{3\kappa^2\eta^4}{8}\Bigg\}  + \nonumber\\
    %&+& \frac{4\pi^2 c^2 e^{\frac{2\pi\Omega c}{a}} \left( e^{\frac{2\pi\Omega c}{a}} +1\right)}{a^2 \left(e^{\frac{2\pi\Omega c}{a}}-1\right)^3} \Bigg\{ 1 + \left( \frac{a^2}{c^2} + \Omega^2\right) \frac{\kappa\eta^2}{6} + \left[ \Omega^2\left( \frac{a^2}{c^2} + \frac{\Omega^2}{5}\right) - \frac{6a^4}{5c^4}\right] \frac{3\kappa^2\eta^4}{4} \Bigg\}\Bigg),
%\end{eqnarray}
\begin{eqnarray}\label{Rp6}
    \overline{\mathcal{R}}^-_{p^6} &\approx& \frac{1}{2\pi}\frac{1}{e^{2\pi/\overline{a}} - 1} \Bigg\{ 1 + \left( 1+ \overline{a}^2 + \frac{3}{\sigma^2} \right)\frac{\overline{\gamma}^2}{6} + \Bigg[ \frac{1}{10} + \frac{\overline{a}^2}{2} + \frac{1}{\sigma^2} + \frac{3\overline{a}^2}{2\sigma^2} - \frac{3\overline{a}^4}{5} \Bigg]\frac{3\overline{\gamma}^4}{4} + \nonumber\\
    &+& \frac{2\pi}{\overline{a}\sigma^2} \frac{e^{2\pi/\overline{a}}}{\left( e^{2\pi/\overline{a}} - 1\right)} \Bigg[ 1 + \left( 1+ \frac{\overline{a}^2}{3}\right) \frac{\overline{\gamma}^2}{2} + \left( 1 + \overline{a}^2 - \frac{3\overline{a}^4}{5}\right)\frac{3\overline{\gamma}^4}{4} + \nonumber\\
    &-& \frac{\pi}{\overline{a}}\frac{\left(e^{2\pi/\overline{a}} + 1\right)}{\left(e^{2\pi/\overline{a}} - 1\right)} \Bigg( 1 + \left( 1+ \overline{a}^2\right) \frac{\overline{\gamma}^2}{6} + \left( \frac{1}{5} + \overline{a}^2 - \frac{6\overline{a}^4}{5} \right) \frac{3\overline{\gamma}^4}{4} \Bigg) \Bigg] \Bigg\},
\end{eqnarray}
\begin{eqnarray}\label{Rp6+}
    \overline{\mathcal{R}}^+_{p^6} &\approx& \frac{1}{2\pi}\frac{e^{2\pi/\overline{a}}}{e^{2\pi/\overline{a}} - 1} \Bigg\{ 1 + \left( 1+ \overline{a}^2 + \frac{3}{\sigma^2} \right)\frac{\overline{\gamma}^2}{6} + \Bigg[ \frac{1}{10} + \frac{\overline{a}^2}{2} + \frac{1}{\sigma^2} + \frac{3\overline{a}^2}{2\sigma^2} - \frac{3\overline{a}^4}{5} \Bigg]\frac{3\overline{\gamma}^4}{4} + \nonumber\\
    &+& \frac{2\pi}{\overline{a}\sigma^2} \frac{1}{\left( e^{2\pi/\overline{a}} - 1\right)} \Bigg[ 1 + \left( 1+ \frac{\overline{a}^2}{3}\right) \frac{\overline{\gamma}^2}{2} + \left( 1 + \overline{a}^2 - \frac{3\overline{a}^4}{5}\right)\frac{3\overline{\gamma}^4}{4} + \nonumber\\
    &-& \frac{\pi}{\overline{a}}\frac{\left(e^{2\pi/\overline{a}} + 1\right)}{\left(e^{2\pi/\overline{a}} - 1\right)} \Bigg( 1 + \left( 1+ \overline{a}^2\right) \frac{\overline{\gamma}^2}{6} + \left( \frac{1}{5} + \overline{a}^2 - \frac{6\overline{a}^4}{5} \right) \frac{3\overline{\gamma}^4}{4} \Bigg) \Bigg] \Bigg\}.
\end{eqnarray}
and for DSR corrections, we have
\begin{eqnarray}\label{Rp4}
    \overline{\mathcal{R}}^{-}_{p^4} &\approx& \frac{1}{2\pi}\frac{1}{e^{2\pi/\overline{a}} - 1} \Bigg\{ 1 +\left( 1 + \overline{a}^2 + \frac{3}{\sigma^2}\right)\frac{\overline{\gamma}^2}{6} + \frac{2\pi}{\overline{a}\sigma^2} \frac{e^{2\pi/\overline{a}}}{\left( e^{2\pi/\overline{a}} - 1\right)} \Bigg[ 1 + \left( 1 + \frac{\overline{a}^2}{3} \right) \frac{\overline{\gamma}^2}{2} +\nonumber\\
    &-& \frac{\pi}{\overline{a}}\frac{\left(e^{2\pi/\overline{a}} + 1\right)}{\left(e^{2\pi/\overline{a}} - 1\right)} \left( 1 + \left( 1 + \overline{a}^2\right) \frac{\overline{\gamma}^2}{6} \right) \Bigg] \Bigg\},
\end{eqnarray}
\begin{eqnarray}\label{Rp4b}
    \overline{\mathcal{R}}^{+}_{p^4} &\approx& \frac{1}{2\pi}\frac{e^{2\pi/\overline{a}}}{e^{2\pi/\overline{a}} - 1} \Bigg\{ 1 +\left( 1 + \overline{a}^2 + \frac{3}{\sigma^2}\right)\frac{\overline{\gamma}^2}{6} + \frac{2\pi}{\overline{a}\sigma^2} \frac{1}{\left( e^{2\pi/\overline{a}} - 1\right)} \Bigg[ 1 + \left( 1 + \frac{\overline{a}^2}{3} \right) \frac{\overline{\gamma}^2}{2} + \nonumber\\
    &-& \frac{\pi}{\overline{a}}\frac{\left(e^{2\pi/\overline{a}} + 1\right)}{\left(e^{2\pi/\overline{a}} - 1\right)} \left( 1 + \left( 1 + \overline{a}^2\right) \frac{\overline{\gamma}^2}{6} \right) \Bigg] \Bigg\}.
\end{eqnarray}

This result shows us that the transition probability rate with high order changes in the dispersion relation also maintains its thermal regime ($\sigma \to \infty$) and that in the limit $\overline{\gamma} \to 0$ we return to the default result (\ref{A3}).

\begin{figure}[H]
    \begin{center}
    \includegraphics[scale=0.42]{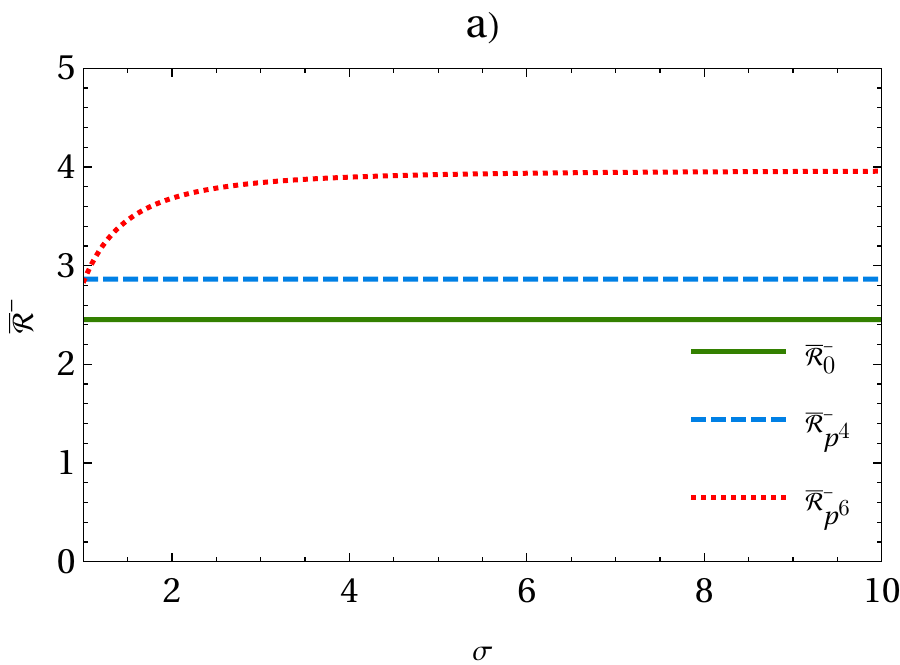}
    \includegraphics[scale=0.42]{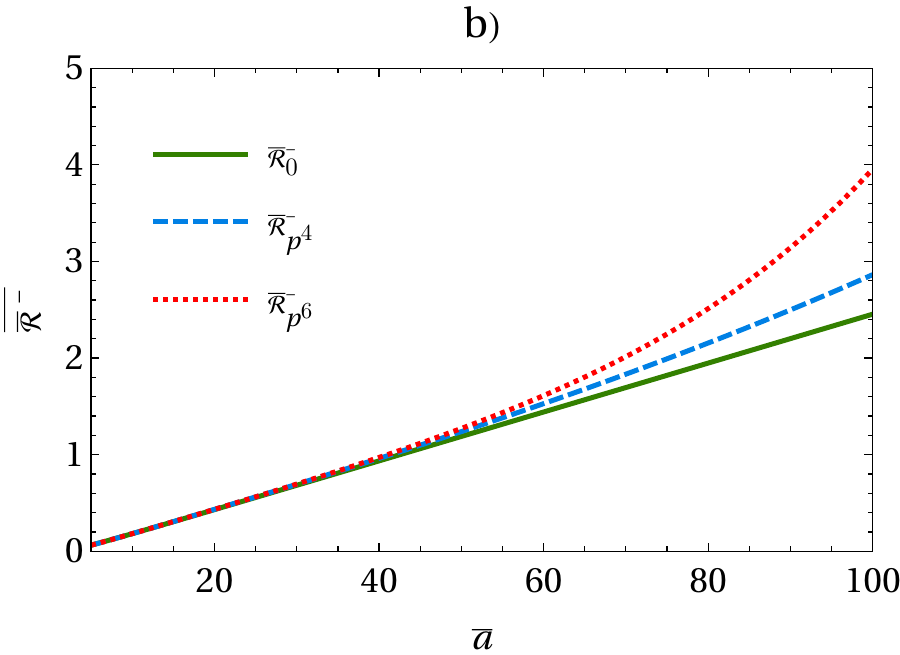}
    \includegraphics[scale=0.42]{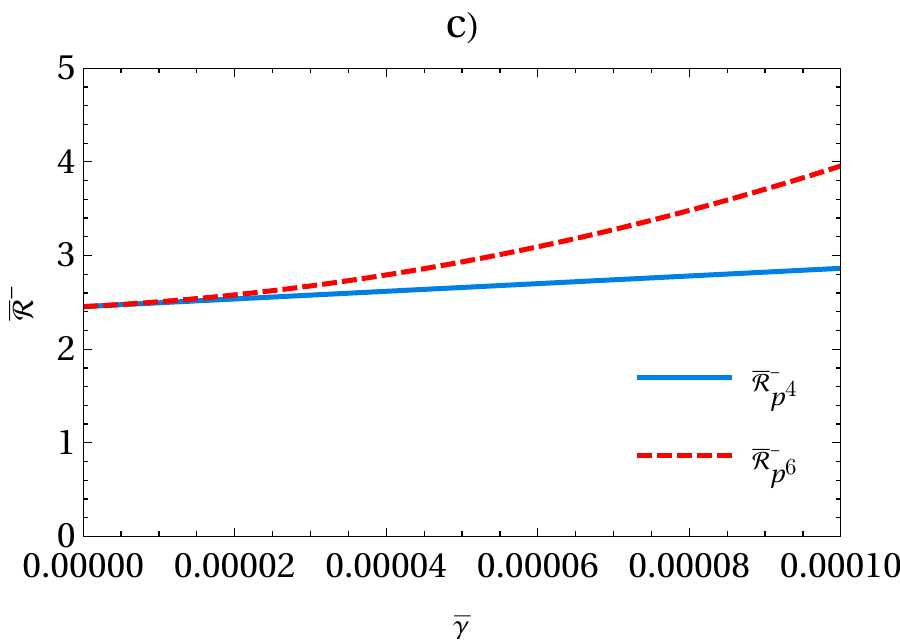}
    \caption{Excitation probability rate $\overline{\mathcal{R}}^-$ as a function of the \textbf{a)} dimensionless parameter $\sigma$, \textbf{b)} dimensionless parameter $\overline{a}$, and \textbf{c)} dimensionless parameter $\overline{\gamma}$. In which the following values were fixed: $\overline{a}=100$, $\sigma=10$, and $\overline{\gamma}=10^{-4}$.}
    \label{Fig 1}
    \end{center}
\end{figure}

The numerical results obtained indicate that the excitation probability rates with modified dispersion relation (with dispersive vacuum) for a detector in an accelerated reference frame behave differently from the case without modifications.
Analyzing excitation probability rates as a function of the dimensionless parameter $\sigma$ (Fig. \ref{Fig 1}a) and as a function of the dimensionless parameter $\overline{a}$ (Fig. \ref{Fig 1}b), we can notice that the cases $\overline{\mathcal{R}}^-_{p^4}$ and $\overline{\mathcal{R}}^-_{p^6}$ are greater than $\overline{\mathcal{R}}^-_{0}$. This result indicates that the detector responds more intensely in cases with dispersive vacuum for the same interaction times and accelerations.
Furthermore, the results on acceleration indicate that around $\overline{a} \approx 50$, with $\overline{\gamma}=10^{-4}$, the effects of the dispersive vacuum begin to appear and then we can visualize the differences between both cases: dispersive and non-dispersive vacuum.
In Fig. \ref{Fig 1}c, we have the excitation probability rates (with changes in the dispersion relation) as a function of the parameter $\overline{\gamma}$, this parameter carries the signature of the dispersive vacuum. This analysis shows that the intensity of the detector's response is greater as the order of modification increases, indicating that the vacuum becomes more dispersive for high-order modifications.
Thus, through these results, we see that changes in the dispersion relation make the vacuum dispersive, and consequently, intensify the detector response.
It is important to highlight that in this work the Unruh effect appears at high accelerations, but some works present proposals in which this effect appears at low accelerations, such as in suitably selected Fock states \cite{Aspachs2010}, detectors that acquire different berry phases \cite{Martinez2011lower}, and cavity optimization \cite{Martinez2021slow, Stargen2022}.

In Fig. \ref{Fig 1}, a value of $\overline{a} = 100$ implies that we are considering a regime of high accelerations compared to the transition frequency in order to observe the Unruh effect significantly, which usually requires high accelerations to be detectable. The high acceleration helps to amplify the detector response, even though it is still on a very small scale. A value of $\sigma = 10$ implies that the interaction time is 10 times greater than the transition time of the two-level system ($\sim \frac{10}{\Omega}$). This means that the detector has enough time to interact with the field, allowing a better measurement of the transition rate. A value of $\overline{\gamma} = 10^{-4}$ suggests that the scale associated with the high-order corrections is very small compared to the energy scale defined by the detector transition frequency. This reflects the fact that spacetime discretization effects are subtle and difficult to detect directly. About the energy ranges required for detection, considering the dispersion relation (\ref{Ep6}) where $\gamma^2 = \kappa\eta^2$ and $\delta^2 = \kappa\eta^4$, if we consider that $\eta$ is of the order of the Planck scale ($\eta \sim 10^{-35}$ m), $\kappa$ can be a factor of order unity, then $\gamma \sim \sqrt{\kappa} \times 10^{-35}$ m and $\delta \sim \sqrt{\kappa} \times 10^{-70}$ m. The energies associated with these scales are very high, far beyond the current capability of direct detection. The transition energy of the $\Omega$ detector must be high enough for the corrections $\gamma^2 p^4$ and $\delta^2 p^6$ to be meaningful. However, in practice this may be beyond the capabilities of current particle accelerators, which operate at much smaller energy scales.

\section{\label{sec:level3}Quantifying the quantum coherence variation}

\begin{figure}[H]
    \centering
    \includegraphics[scale=0.44]{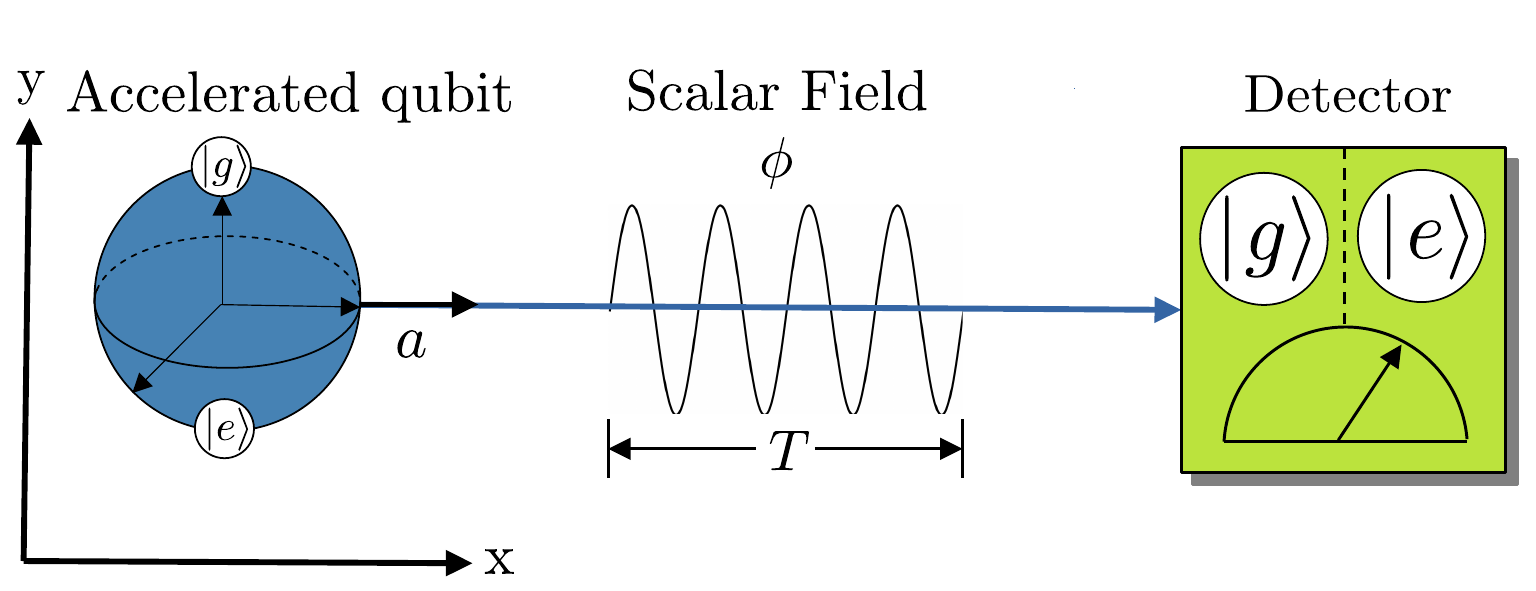}
    \caption{Illustration of a quantum scheme for estimating the variation of coherence of a single qubit induced by the Unruh effect. A detector, initially prepared in the superposition state $|\psi_{\mathrm{D}}\rangle$, undergoes uniform acceleration, interacts with a massless scalar field $\phi$ for a finite time interval $T$ and subsequently has its internal states measured.}
    \label{Fig 2}
\end{figure}

Quantum computation and information, are built upon the notion of the quantum bit, or \textit{qubit}. Qubits are essentially realized as physical systems; yet, we predominantly view them as abstract mathematical entities. This abstraction affords us the freedom to formulate a general theory independent of specific systems for implementation \cite{nielsen2010}. In our setup, as shown in Fig. \ref{Fig 2}, there is a detector and a free massless scalar field. The field is initially in the vacuum state, while the detector is prepared in a superposition state $|\psi_{\mathrm{D}}\rangle = \alpha|g\rangle + \beta|e\rangle$, representing a qubit state with $|\alpha|^2 + |\beta|^2 = 1$. To express $|\psi_{\mathrm{D}}\rangle$ more conveniently, we represent it as a Bloch vector:
\begin{eqnarray}
|\psi_{\mathrm{D}}\rangle = \cos\frac{\theta}{2}e^{i\chi/2}|g\rangle + \sin\frac{\theta}{2}e^{-i\chi/2}|e\rangle,
\end{eqnarray}
where $\alpha = \cos{\frac{\theta}{2}} e^{i\frac{\chi}{2}}$ and $\beta = \sin{\frac{\theta}{2}} e^{-i\frac{\chi}{2}}$ and with density matrix $\hat{\rho}^{\mathrm{in}}_D = |\psi_D\rangle \langle\psi_D|$ which becomes
\begin{eqnarray}
    \hat{\rho}^{\mathrm{in}}_D = |\alpha|^2 |g\rangle\langle g| + \alpha\beta^*|g\rangle\langle e| + \alpha^*\beta|e\rangle\langle g| + |\beta|^2|e\rangle\langle e|. \label{rhoinD}
\end{eqnarray}
Here, $\theta \in [0,\pi]$ and $\chi \in [0,2\pi]$ are the polar and azimuthal angles in the Bloch sphere, respectively. After its preparation, the detector undergoes uniform acceleration in a linear accelerator, followed by interaction with the scalar field $\phi$ for a specific duration denoted as $T$. Finally, the internal states of the detector are measured.
Let us assume the initial state of our system accelerated detector and the scalar field. We proceed with the assumption that the preparation of this system adheres to a defined structure: $\hat{\rho}_{\mathrm{in}} = \hat{\rho}^{\mathrm{in}}_D \otimes \hat{\rho}_\phi$, where $\hat{\rho}_\phi = |0\rangle \langle 0|$. In this context, the term $|0\rangle$ is the vacuum state (dispersive or non-dispersive) of the field. To understand the dynamics of this system we must first investigate the detector-field interaction. Guided by (\ref{Hint}) with weak coupling ($g$ being small), we can obtain the final density matrix described by
\begin{eqnarray}
    \hat{\rho}_{\mathrm{out}} &=& \hat{\rho}_{\mathrm{out}}^{(0)} + \hat{\rho}_{\mathrm{out}}^{(1)} + \hat{\rho}_{\mathrm{out}}^{(2)} + \mathcal{O}(g^3),
\end{eqnarray}
which carries with itself unitary transformations, so that
\begin{eqnarray}
    \hat{\rho}_{\mathrm{out}}^{(0)} &=& \mathcal{\hat{U}}^{(0)} \hat{\rho}_{\mathrm{in}} \mathcal{\hat{U}}^{(0)^\dagger},\\
    \hat{\rho}_{\mathrm{out}}^{(1)} &=& \mathcal{\hat{U}}^{(1)} \hat{\rho}_{\mathrm{in}} + \hat{\rho}^{\mathrm{in}}\mathcal{\hat{U}}^{(0)^\dagger},\\
    \hat{\rho}_{\mathrm{out}}^{(2)} &=& \mathcal{\hat{U}}^{(1)} \hat{\rho}_{\mathrm{in}} \mathcal{\hat{U}}^{(1)^\dagger} + \mathcal{\hat{U}}^{(2)} \hat{\rho}_{\mathrm{in}} \mathcal{\hat{U}}^{(2)^\dagger}.
\end{eqnarray}
The temporal evolution operator is described, in a perturbative way, by
\begin{eqnarray}
    \mathcal{\hat{U}} = \mathcal{\hat{U}}^{(0)} + \mathcal{\hat{U}}^{(1)} + \mathcal{\hat{U}}^{(2)} + \mathcal{O}(g^3)
\end{eqnarray}
where
\begin{eqnarray}
    \mathcal{\hat{U}}^{(0)} &=& \mathbb{I},\\
    \mathcal{\hat{U}}^{(1)} &=& -ig\int_{-\infty}^{\infty}d\tau\Theta(\tau)\mu(\tau)\phi[x(\tau)],\\
    \mathcal{\hat{U}}^{(2)} &=& -g^2\int^{+\infty}_{-\infty} d\tau \int^{+\tau}_{-\infty} d\tau' \Theta(\tau)\Theta(\tau') \mu(\tau)\mu(\tau') \phi[x(\tau)]\phi[x(\tau')],
\end{eqnarray}
where the monopole moment of qubit can be written as $\mu(\tau) = [\hat{\sigma}_{+}e^{i\Omega\tau} + \hat{\sigma}_{-}e^{-i\Omega\tau}]$, and $\hat{\sigma}_{+} = |e\rangle\langle g|$ and $\hat{\sigma}_{-} = |g\rangle\langle e|$ are the raising and lowering operators, respectively.
When our focus narrows to the observation of the detector's state, excluding consideration of the field, a distinct revelation unfolds: the detector manifests itself in either the excited state or the ground state. Nevertheless, this discernible state no longer adheres to the purity of its initial form. The resultant condition is aptly captured by the reduced density matrix, denoted as $\hat{\rho}^{\mathrm{out}}_D = \mathrm{Tr}_{|0\rangle}[\hat{\rho}_{\mathrm{out}}]$. The reduced density matrix for the detector finds its definition in the following manner:
\begin{eqnarray}
    \hat{\rho}^{\mathrm{out}}_D = \hat{\rho}_{\mathrm{in}}^{D} + \mathrm{Tr}_{|0\rangle} \left(\mathcal{\hat{U}}^{(1)}\hat{\rho}_{\mathrm{in}}\mathcal{\hat{U}}^{(1)\dagger}\right) + \mathrm{Tr}_{|0\rangle} \left(\mathcal{\hat{U}}^{(2)}\hat{\rho}_{\mathrm{in}}\right) + \mathrm{Tr}_{|0\rangle} \left(\hat{\rho}_{\mathrm{in}}\mathcal{\hat{U}}^{(2)\dagger}\right),
    \label{rhooutD}
\end{eqnarray}
and after extensive and tedious calculations, we obtain that
\begin{eqnarray}
\mathrm{Tr}_{|0\rangle} \left(\mathcal{\hat{U}}^{(1)}\hat{\rho}_{\mathrm{in}}\mathcal{\hat{U}}^{(1)\dagger}\right) &=& g^2 \int^{+\infty}_{-\infty} d\tau \int^{+\infty}_{-\infty} d\tau' \Theta(\tau)\Theta(\tau') D(\tau, \tau') \times\nonumber\\
    &\times& \Big[\alpha\beta^* e^{i\Omega(\tau+\tau')}|e\rangle\langle g| + |\alpha|^2 e^{-i\Omega(\tau-\tau')}|e\rangle\langle e| + \nonumber\\
    &+& |\beta|^2 e^{-i\Omega(\tau-\tau')}|g\rangle\langle g| + \beta\alpha^* e^{-i\Omega(\tau+\tau')}|g\rangle\langle e|\Big],\label{term1}\\
\mathrm{Tr}_{|0\rangle} \left(\mathcal{\hat{U}}^{(2)}\hat{\rho}_{\mathrm{in}}\right) &=& -g^2 \int^{+\infty}_{-\infty} d\tau \int^{+\tau}_{-\infty} d\tau_1 \Theta(\tau)\Theta(\tau_1) D(\tau, \tau_1) \nonumber\\
    &\times& \Big[e^{i\Omega(\tau-\tau_1)} \left( \beta\alpha^*|e\rangle\langle g| + |\beta|^2|e\rangle\langle e| \right) + \nonumber\\
    &+& e^{-i\Omega(\tau-\tau_1)} \left( |\alpha|^2|g\rangle\langle g| + \alpha\beta^*|g\rangle\langle e| \right) \Big],\label{term2}\\
\mathrm{Tr}_{|0\rangle} \left(\hat{\rho}_{\mathrm{in}}\mathcal{\hat{U}}^{(2)\dagger}\right) &=& -g^2 \int^{+\infty}_{-\infty} d\tau \int^{+\tau}_{-\infty} d\tau_1 \Theta(\tau)\Theta(\tau_1) D^*(\tau, \tau_1) \nonumber\\
    &\times& \Big[e^{-i\Omega(\tau-\tau_1)} \left( \alpha\beta^*|g\rangle\langle e| + |\beta|^2|e\rangle\langle e| \right) + \nonumber\\
    &+& e^{i\Omega(\tau-\tau_1)} \left( |\alpha|^2|g\rangle\langle g| + \beta\alpha^*|e\rangle\langle g| \right)\Big].\label{term3}
\end{eqnarray}
Therefore, defining the integrals as follows
\begin{eqnarray}
    \mathcal{C}^{\pm}(\Omega) &=& \int^{+\infty}_{-\infty} d\tau \int^{+\infty}_{-\infty} d\tau' \Theta(\tau)\Theta(\tau') e^{\pm i\Omega(\tau+\tau')} D(\tau, \tau'),\label{C+-} \\
    \mathcal{G}^{\pm}(\Omega) &=& \int^{+\infty}_{-\infty} d\tau \int^{+\tau}_{-\infty} d\tau' \Theta(\tau)\Theta(\tau') e^{\pm i\Omega(\tau-\tau')} D(\tau, \tau'),\label{intG}
\end{eqnarray}
and substituting (\ref{intF}), (\ref{rhoinD}), (\ref{term1})-(\ref{intG}) into (\ref{rhooutD}), we have
\begin{eqnarray}\label{rhoD}
    \rho^{\mathrm{out}}_D &=& \left\{ \cos^2{\frac{\theta}{2}} + g^2\left[ \sin^2{\frac{\theta}{2}} \mathcal{F}^- - 2\cos^2{\frac{\theta}{2}} \mathbf{Re}\left(\mathcal{G}^-\right) \right] \right\} |g\rangle \langle g| \nonumber\\
    &+& \left\{ \cos{\frac{\theta}{2}}\sin{\frac{\theta}{2}}e^{-i\chi} + g^2\left[ \cos{\frac{\theta}{2}}\sin{\frac{\theta}{2}} \left((e^{i\chi}\mathcal{C}^+ - e^{-i\chi} \left(\mathcal{G}^+ + \mathcal{G}^{-*}\right) \right) \right] \right\} |e\rangle \langle g| \nonumber\\
    &+& \left\{ \cos{\frac{\theta}{2}}\sin{\frac{\theta}{2}}e^{i\chi} + g^2\left[ \cos{\frac{\theta}{2}}\sin{\frac{\theta}{2}} \left((e^{-i\chi}\mathcal{C}^- - e^{i\chi} \left(\mathcal{G}^- + \mathcal{G}^{+*}\right) \right) \right] \right\} |g\rangle \langle e| \nonumber\\
    &+& \left\{ \sin^2{\frac{\theta}{2}} + g^2\left[ \cos^2{\frac{\theta}{2}} \mathcal{F}^+ - 2\sin^2{\frac{\theta}{2}} \mathbf{Re}\left(\mathcal{G}^+\right) \right] \right\} |e\rangle \langle e|,
\end{eqnarray}
where here in this setup $\mathbf{Re}(\mathcal{G}^-) = \mathbf{Re}(\mathcal{G}^+)$, thus note that,
\begin{eqnarray}
    \mathbf{Re}(\mathcal{G}^-) = \frac{1}{2} \left( \mathcal{F}^-\sin^2{\frac{\theta}{2}} + \mathcal{F}^+\cos^2{\frac{\theta}{2}}\right),
    \label{RelationTraceless}
\end{eqnarray}
with $\mathcal{F}^\pm = \sigma\mathcal{R}^\pm$, and $\rho^{\mathrm{out}}_D$ is traceless in $g^2$ \cite{Martinez2014zeromode, Martinez2014signaling, Martinez2015Causality}. Furthermore, we have to solve the integrals that appeared, for the case without modification in the dispersion relation the Eq. (\ref{C+-}) becomes
\begin{equation}\label{C0}
    \mathcal{C}^{\pm}_0 = \frac{\overline{a}\sigma\sqrt{\pi}}{4\pi^2} e^{-\sigma^2},
\end{equation}
for more details see \cite{lopes2021}. The solutions of Eq. (\ref{C+-}) for the cases with modified dispersion relation were obtained in detail in the Appendix \ref{appendixA} after extensive calculations, thus, these solutions are given by
\begin{eqnarray}\label{Cp4}
    \mathcal{C}^\pm_{p^4} &=& \frac{\overline{a}\sigma\sqrt{\pi}}{4\pi^2} e^{-\sigma^2} \Bigg[ 1 + \left( \overline{a}^2 + \frac{1}{2\sigma^2}\right) \frac{\overline{\gamma}^2}{6}\Bigg],
\end{eqnarray}
and
\begin{eqnarray}\label{Cp6}
    \mathcal{C}^\pm_{p^6} &=& \frac{\overline{a}\sigma\sqrt{\pi}}{4\pi^2} e^{-\sigma^2} \Bigg[ 1 + \left( \overline{a}^2 + \frac{1}{2\sigma^2}\right) \frac{\overline{\gamma}^2}{6} + \left( \frac{\overline{a}^2}{3\sigma^2} + \frac{1}{20\sigma^4} - \frac{4\overline{a}^4}{5}\right) \frac{9\overline{\gamma}^4}{16}\Bigg].
\end{eqnarray}

\begin{figure}[H]
    \centering
    \includegraphics[scale=0.85]{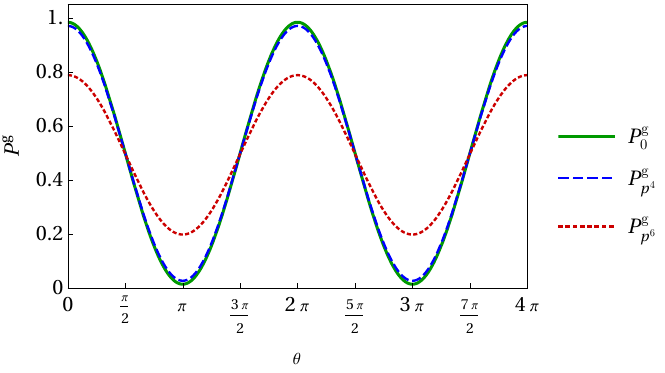}
    \caption{Probabilities $P^{\mathrm{g}}$ as a function of the polar angle $\theta$ of the Bloch sphere, in which the following values were fixed: $\overline{a} = 100$, $g = 0.025$, $\sigma = 10$, and $\overline{\gamma} = 5 \times 10^{-4}$.}
    \label{Fig 3}
\end{figure}

We know that obtaining probability in quantum measurements is a fundamental aspect for understanding, predicting, and exploring quantum phenomena. It is worth noting that by employing the equation (\ref{rhoD}), we can compute the probability associated with detecting the system in the state $|g\rangle$, is given $P^{\mathrm{g}} = \langle g|\rho_D^{\mathrm{out}}|g\rangle$, so we get
\begin{align}\label{Pg}
    P^{\mathrm{g}} =  \cos^2\frac{\theta}{2} + \sigma g^2\left( \overline{\mathcal{R}}^- \sin^4\frac{\theta}{2} - \overline{\mathcal{R}}^+ \cos^4\frac{\theta}{2} \right),
\end{align}
where we use the relationship (\ref{RelationTraceless}). It is possible to observe that (\ref{Pg}) depends on the transition probability rate, so to obtain the probability without or with a modified dispersion relation, simply substitute the corresponding transition probability rate. Precisely, to obtain $P^{\mathrm{g}}_{0}$ one uses (\ref{A3}) and (\ref{A3b}), for $P^{\mathrm{g}}_{p^4}$ one uses (\ref{Rp4}) and (\ref{Rp4b}) and, finally, to obtain $P^{\mathrm{g}}_{p^6}$ we use (\ref{Rp6}) and (\ref{Rp6+}). In Fig. \ref{Fig 3}, we plot the probability as a function of the polar angle $\theta$. The results reveal a significant reduction in the amplitude of probability oscillation in cases where dispersive vacuum, with this effect being more pronounced for $P^{\mathrm{g}}_{p^6}$ compared to $P^{\mathrm{g}}_{p^4}$, considering $\overline{\gamma} = 5 \times 10^{-4}$. Here we once again have a result that indicates that the vacuum becomes more dispersive as the order of modification increases.

Within the framework of a quantum system consisting of two levels, the term $l^1$ norm coherence commonly denotes the degree of coherence existing between the system's two fundamental basis states, typically denoted as $\left|g\right\rangle$ and $\left|e\right\rangle$. This coherence essentially captures the quantum phenomena of superposition and interference manifested between these states, as extensively discussed in literature \cite{Streltsov2017, wu2020}. In a mathematical context, specifically for a two-level quantum system, the $l^1$ norm coherence is determined by the summation of the absolute values of the off-diagonal elements present in the density matrix $\hat{\rho}$ governing the description of the system \cite{Baumgratz2014QuantifyingCoherence}, is written as 
\begin{align}\label{Ql1}
    \mathcal{Q}^{l^1}(\hat{\rho}) = \sum_{i \neq j} \mid \hat{\rho}^{ij}\mid,
\end{align}
and in our case it becomes
\begin{eqnarray}
    \mathcal{Q}^{l^1}(\hat{\rho}_D^{\mathrm{out}}) =  |\hat{\rho}_D^{(\mathrm{out})\mathrm{ge}}| + |\hat{\rho}_D^{(\mathrm{out})\mathrm{eg}}|,
    \label{Q1}
\end{eqnarray}
where
\begin{eqnarray}
    \hat{\rho}_D^{(\mathrm{out})\mathrm{ge}} &=& \cos{\frac{\theta}{2}}\sin{\frac{\theta}{2}}e^{i\chi} + g^2\left[ \cos{\frac{\theta}{2}}\sin{\frac{\theta}{2}} \left((e^{-i\chi}\mathcal{C}^- - e^{i\chi} \left(\mathcal{G}^- + \mathcal{G}^{+*}\right) \right) \right], \nonumber\\
    \hat{\rho}_D^{(\mathrm{out})\mathrm{eg}} &=& \cos{\frac{\theta}{2}}\sin{\frac{\theta}{2}}e^{-i\chi} + g^2\left[ \cos{\frac{\theta}{2}}\sin{\frac{\theta}{2}} \left((e^{i\chi}\mathcal{C}^+ - e^{-i\chi} \left(\mathcal{G}^+ + \mathcal{G}^{-*}\right) \right) \right],
\end{eqnarray}
and note that Eq. (\ref{Q1}) can be written as
\begin{eqnarray}
    \mathcal{Q}^{l^1}(\hat{\rho}_D^{\mathrm{out}}) =  2\left[ \left( \hat{\rho}_D^{(\mathrm{out})\mathrm{ge}} \right) \left(\hat{\rho}_D^{(\mathrm{out})\mathrm{ge}}\right)^*\right]^{\frac{1}{2}}
\end{eqnarray}
and at last we have
\begin{eqnarray}
    \mathcal{Q}^{l^1}(\hat{\rho}_D^{\mathrm{out}}) =  \sin{\theta} \Big\{ 1 - g^2\left[ 2\mathbf{Re}\left(\mathcal{G}^-\right) - \mathcal{C}^- \cos{2\chi}\right] \Big\} + \mathcal{O}(g^4),
    \label{Q2}
\end{eqnarray}
where we use the following approximation $(1+x)^{-n} \approx 1 - nx + \cdots$. Quantum coherence between internal states is lost
 gradually due to the thermal environment created through acceleration. Now, we can obtain the coherence variation by investigating how much the initial coherence $\mathcal{Q}^{l^1}(\hat{\rho}_D^{\mathrm{in}})$ varied compared to the final coherence $\mathcal{Q}^{l^1}(\hat{\rho}_D^{\mathrm{out}})$ after interacting with the massless scalar field. Therefore, the quantum coherence variation is defined as
\begin{eqnarray}
    \Delta\mathcal{Q}^{l^1} \equiv \mathcal{Q}^{l^1}(\hat{\rho}_D^{\mathrm{in}}) - \mathcal{Q}^{l^1}(\hat{\rho}_D^{\mathrm{out}}),
\end{eqnarray}
where $\mathcal{Q}^{l^1}(\hat{\rho}_D^{\mathrm{in}}) = \sin{\theta}$, is the $l^1$ norm coherence of a qubit state described by the density matrix given by Eq. (\ref{rhoinD}), and we get
\begin{eqnarray}
    \Delta\mathcal{Q}^{l^1} = \sin{\theta} \Big\{g^2\left[ 2\mathbf{Re}\left(\mathcal{G}^-\right) - \mathcal{C}^- \cos{2\chi}\right] \Big\} + \mathcal{O}(g^4).
    \label{DeltaQ}
\end{eqnarray}
We obtain the coherence variation per unit $g^2$ for the case without and with modified dispersion relation. First, let us consider the case without modified dispersion relation. By substituting the expressions (\ref{RelationTraceless}) and (\ref{C0}) into (\ref{DeltaQ}), we find
\begin{eqnarray}\label{DeltaQl0}
\overline{\Delta\mathcal{Q}}^{l^1}_0 =  \sigma \sin{\theta}\left[ \overline{\mathcal{R}}_0^- \sin^2{\frac{\theta}{2}} + \overline{\mathcal{R}}_0^+ \cos^2{\frac{\theta}{2}} - \frac{\overline{a}\sqrt{\pi}}{4\pi^2}e^{-\sigma^2} \cos{2\chi}\right] + \mathcal{O}(g^4),
\end{eqnarray}
where $\overline{\Delta\mathcal{Q}}^{l^1} = \Delta\mathcal{Q}^{l^1}/g^2$. By doing the same procedure, we can obtain the corresponding expressions for the cases with modified dispersion relation, and then we have to replace Eq. (\ref{RelationTraceless}) and (\ref{Cp4}) into (\ref{DeltaQ}), so we have
\begin{eqnarray}\label{Ql4}
\overline{\Delta\mathcal{Q}}^{l^1}_{p^4} &=& \sigma \sin{\theta}\Bigg\{ \overline{\mathcal{R}}_{p^4}^- \sin^2{\frac{\theta}{2}} + \overline{\mathcal{R}}_{p^4}^+ \cos^2{\frac{\theta}{2}} +\nonumber\\
&-& \frac{\overline{a}\sqrt{\pi}}{4\pi^2}e^{-\sigma^2} \left[ 1 + \left( \overline{a}^2 + \frac{1}{2\sigma^2}\right)\frac{\overline{\gamma}^2}{6} \right] \cos{2\chi}\Bigg\} + \mathcal{O}(g^4),
\end{eqnarray}
and similarly we substitute Eq. (\ref{RelationTraceless}) and (\ref{Cp6}) into (\ref{DeltaQ}), and finally results in
\begin{eqnarray}\label{Ql6}
\overline{\Delta\mathcal{Q}}^{l^1}_{p^6} &=& \sigma \sin{\theta} \Bigg\{ \overline{\mathcal{R}}_{p^6}^- \sin^2{\frac{\theta}{2}} + \overline{\mathcal{R}}_{p^6}^+ \cos^2{\frac{\theta}{2}} - \frac{\overline{a}\sqrt{\pi}}{4\pi^2}e^{-\sigma^2} \Bigg[ 1 + \left( \overline{a}^2 + \frac{1}{2\sigma^2}\right)\frac{\overline{\gamma}^2}{6} +\nonumber\\
&+& \left( \frac{\overline{a}^2}{3\sigma^2} + \frac{1}{20\sigma^4} - \frac{4\overline{a}^4}{5}\right)\frac{9\overline{\gamma}^4}{16} \Bigg] \cos{2\chi}\Bigg\} + \mathcal{O}(g^4).
\end{eqnarray}

To ensure the validity of the perturbation theory, the parameter $\overline{\gamma}$ must be small, so one must have $\Omega\gamma \ll 1$, which implies that the transition frequency $\Omega$ and the correction $\gamma$ (related to the characteristic length scale of spacetime $\eta$) must be small compared to unity. Furthermore, the coupling constant $g$, which measures the strength of the interaction between the detector and the field, must be sufficiently small ($g \ll 1$) to ensure the validity of the perturbation theory. These conditions ensure that the corrections due to high-order terms in the dispersion relation are small and that the perturbative development is consistent.

\begin{figure}[H]
    \begin{center}
    \includegraphics[scale=0.46]{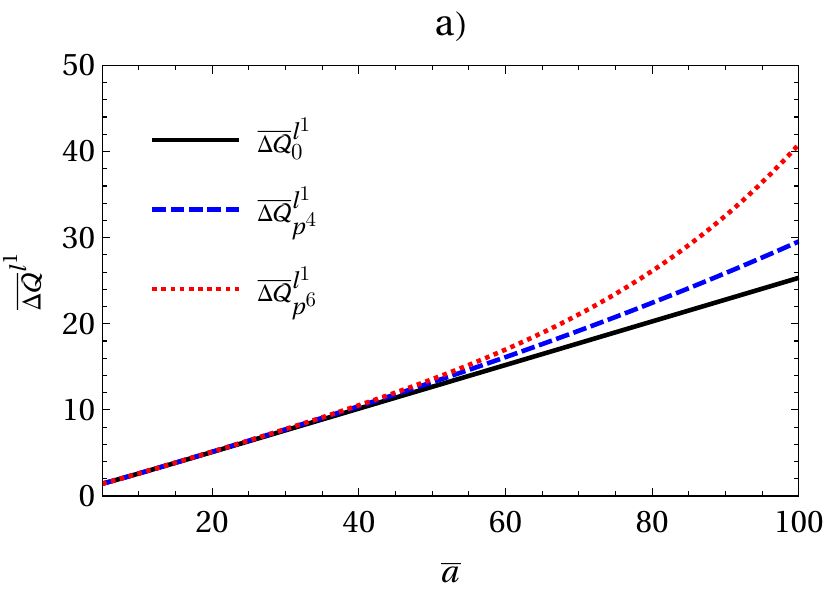}
    \includegraphics[scale=0.46]{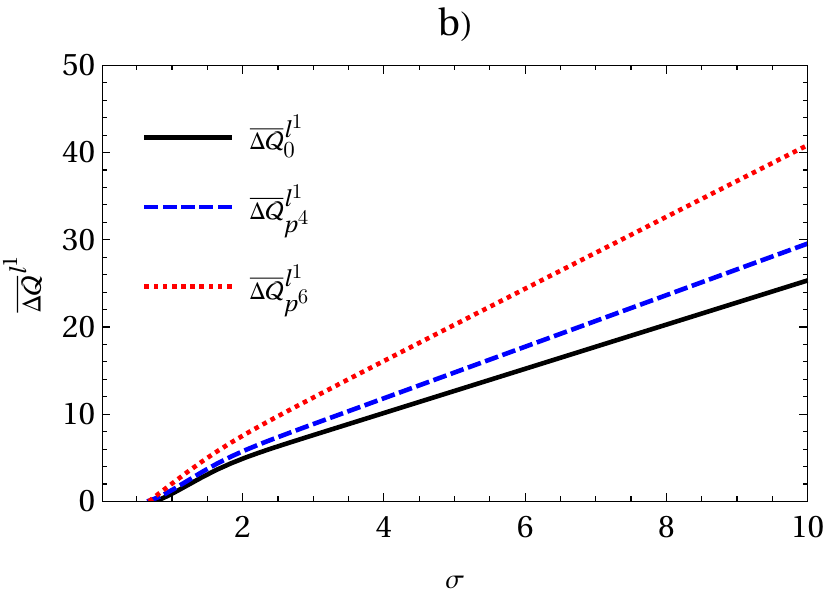}
    \includegraphics[scale=0.46]{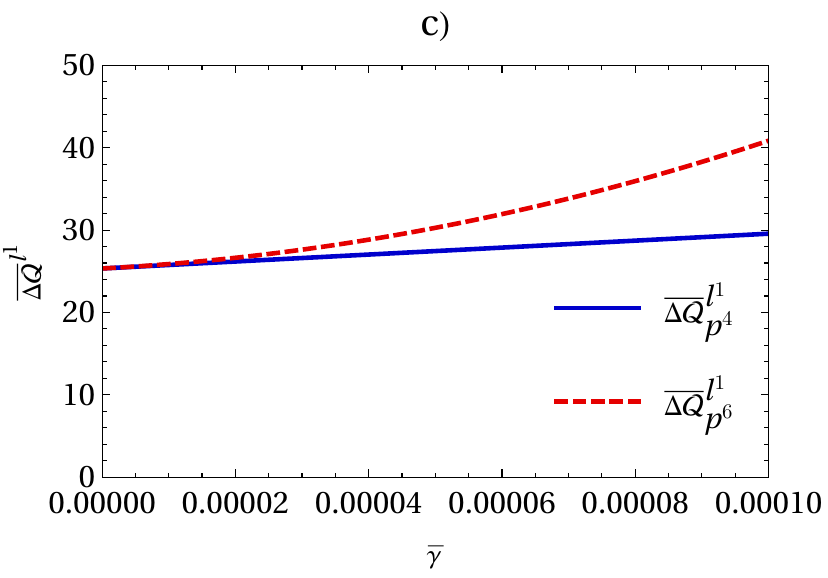}\\
    \caption{$\overline{\Delta\mathcal{Q}}^{l^1}$ \textbf{a)} as a function of the dimensionless parameter $\overline{a}$, \textbf{b)} as a function of the dimensionless parameter $\sigma$, and \textbf{c)} as a function of the dimensionless parameter $\overline{\gamma}$. In which the following values were fixed: $\theta=\pi/2$, $\chi=0$, $\overline{a}=100$, and $\sigma=10$, $\overline{\gamma} = 10^{-4}$.}
    \label{fig 4}
    \end{center}
\end{figure}

We perform a numerical analysis of the coherence variation as a function of the dimensionless parameter $\overline{a}$ (Fig. \ref{fig 4}a) and as a function of the dimensionless parameter $\sigma$ (Fig. \ref{fig 4}b). This plot shows that the decoherence process is greater for cases with modified dispersion relations when compared to the case without modifications, for the same interaction times and accelerations. Once again analyzing the acceleration, it can be seen that the effects of the dispersive vacuum begin to appear around $\overline{a} \approx 50$ considering the dimensionless parameter $\overline{\gamma} = 10^{-4}$.
In Fig. \ref{fig 4}c, we plot the variation of coherence with dispersive vacuum as a function of the parameter $\overline{\gamma}$, we note that $\overline{\Delta\mathcal{Q}}^{l^1}_{p^6}$ varies more quickly than $\overline{\Delta\mathcal{Q}}^{l^1}_{p^4}$. This analysis indicates that the variation in coherence increases according to the order of modification, this is due to the vacuum becoming more dispersive as we change the dispersion relation.
The modifications used here give the vacuum an index of refraction that intensifies the quantum fluctuations that are captured by detector-field interaction. Therefore, these changes in quantum fluctuations transfer energy to the particles observed in the non-inertial reference frame and consequently intensify the environmental effect that is responsible for the decoherence process. 

\section{Conclusions}

We investigated the effects caused by dispersive vacuum on the decoherence of an Unruh-DeWitt detector. Modifications in the dispersion relation give the vacuum a refractive index making it dispersive. Our findings showed that the dispersive vacuum adds significant effects to the detector response and the decoherence mechanism. The results obtained on the transition probability rates showed that the dispersive vacuum intensifies the detector response. When analyzing the probability of finding the system in the ground state as a function of the polar angle of the Bloch sphere, we observed that the dispersive vacuum amplifies the effects of decreasing the probability. Additionally, when studying the behavior of the variation of quantum coherence for the cases with dispersive and non-dispersive vacuum, the results show that the consideration of a dispersive vacuum amplifies the decoherence process of the system considered here. Vacuum dispersive intensifies the quantum fluctuations that are captured by the detector-field interaction. On the other hand, these changes in the quantum fluctuations intensify the creation of particles due to the Unruh effect that increase the ambient effect, and the result of this is the amplification of the process of decoherence of the two-level system. Therefore, we can say that the dispersive vacuum plays the role of an amplifier of decoherence.

\backmatter

%\bmhead{Supplementary information}

%If your article has accompanying supplementary file/s please state so here. 

%Authors reporting data from electrophoretic gels and blots should supply the full unprocessed scans for key as part of their Supplementary information. This may be requested by the editorial team/s if it is missing.

%Please refer to Journal-level guidance for any specific requirements.

\bmhead{Acknowledgements}

The authors express their gratitude to Robert H. Jonsson and Eduardo Martín-Martínez for useful discussion and critical comments. PHMB acknowledges the Brazilian funding agency CAPES for financial support.

\section*{Declarations}

Not applicable.

\section*{Data Availability Statement}

No Data associated in the manuscript.

\begin{appendices}

\section{\label{appendixA}Calculation of $\mathcal{C}^\pm_{p^4}$ and $\mathcal{C}^\pm_{p^6}$}

Here in this appendix, we obtain, in detail, $\mathcal{C}^\pm$ with the modified dispersion relation for DSR and HLG. We first calculate $\mathcal{C}^\pm_{p^6}$ and then we will apply the limit $\delta \to 0$ to obtain $\mathcal{C}^\pm_{p^4}$, so we have
\begin{eqnarray}\label{intCp6}
    \mathcal{C}^\pm_{p^6} = \int^{+\infty}_{-\infty}d\tau \int^{+\infty}_{-\infty}d\tau' \Theta(\tau) \Theta(\tau') e^{\pm i\Omega(\tau+\tau')} D^+_{p^6}(x,x'),
\end{eqnarray}
substituting the Eq. (\ref{D+p6}) into Eq. (\ref{intCp6}),
\begin{eqnarray}
    \mathcal{C}^\pm_{p^6} &=& \int^{+\infty}_{-\infty}d\tau \int^{+\infty}_{-\infty}d\tau' \Theta(\tau) \Theta(\tau') e^{\pm i\Omega(\tau+\tau')}  \Bigg[ \frac{-1}{4\pi^2} \sum^{\infty}_{k=-\infty} \frac{\left(1 + \frac{a^2\gamma^2}{6} - \frac{3a^4\delta'^2}{5}\right)}{\left( \tau-\tau' - 2i\epsilon - 2\pi ik/a\right)^{2}} + \nonumber\\
    &+& \frac{1}{4\pi^2} \sum^{\infty}_{k=-\infty} \frac{\left(\gamma^2 + 3a^2\delta'^2\right)}{\left( \tau-\tau' - 2i\epsilon - 2\pi ik/a\right)^{4}} - \frac{1}{4\pi^2} \sum^{\infty}_{k=-\infty} \frac{12\delta'^2}{\left( \tau-\tau' - 2i\epsilon - 2\pi ik/a\right)^{6}}\Bigg].\nonumber\\
\end{eqnarray}
To compact the calculations we can make the following definitions
\begin{eqnarray}
    A &=& 1 + \frac{a^2\gamma^2}{6} - \frac{3a^4\delta'^2}{5}, \\
    B &=& \gamma^2 + 3a^2\delta'^2,\\
    C &=& 12\delta'^2,
\end{eqnarray}
and through these definitions we have
\begin{eqnarray}
    \mathcal{C}^\pm_{p^6} &=& \int^{+\infty}_{-\infty}d\tau \int^{+\infty}_{-\infty}d\tau' \Theta(\tau) \Theta(\tau') e^{\pm i\Omega(\tau+\tau')} \Bigg[ \frac{-1}{4\pi^2} \sum^{\infty}_{k=-\infty} \frac{A}{\left( \tau-\tau' - 2i\epsilon - 2\pi ik/a\right)^{2}} + \nonumber\\
    &+& \frac{1}{4\pi^2} \sum^{\infty}_{k=-\infty} \frac{B}{\left( \tau-\tau' - 2i\epsilon - 2\pi ik/a\right)^{4}} - \frac{1}{4\pi^2} \sum^{\infty}_{k=-\infty} \frac{C}{\left( \tau-\tau' - 2i\epsilon - 2\pi ik/a\right)^{6}}\Bigg],\nonumber\\
\end{eqnarray}
and making the variable substitutions $s=\tau-\tau'$ and $u=\tau+\tau'$, replacing the expression of the window function, separating the integrations of $s$ and $u$, completing the square of exponential of $u$ i.e. $e^{\pm i\Omega u}e^{-u^2/4T^2} = e^{-(u/2T \mp iT\Omega)^2}e^{-T^2\Omega^2}$, and we obtain the following expression (after integrating into $u$), so it is obvious that the integral in $u$ is Gaussian and equal to $\sqrt{\pi}$, 
\begin{eqnarray}
    \mathcal{C}^\pm_{p^6} &=& -\frac{\sqrt{\pi}}{8\pi^2} e^{-T^2\Omega^2}\sum^{\infty}_{k=-\infty} \int^{+\infty}_{-\infty} e^{\frac{-s^2}{4T^2}}ds \times \nonumber\\
    &\times&  \Bigg[ \frac{A}{\left( s - 2i\epsilon - 2\pi ik/a\right)^{2}} - \frac{B}{\left( s - 2i\epsilon - 2\pi ik/a\right)^{4}} + \frac{C}{\left( s - 2i\epsilon - 2\pi ik/a\right)^{6}}\Bigg],
\end{eqnarray}
and changing the variable as follows $y=2T$, writing $e^{-y^2}$ as a Fourier transform of the form $e^{-y^2} = \frac{1}{\sqrt{\pi}}\int^{+\infty}_{-\infty}d\omega e^{-\omega^2}e^{2iyk}$, we obtain 
\begin{eqnarray}\label{cp6sem_ints}
    \mathcal{C}^\pm_{p^6} &=& -\frac{e^{-T^2\Omega^2}}{8\pi^2} \sum^{\infty}_{k=-\infty} \int^{+\infty}_{-\infty} e^{-\omega^2}e^{2iyk}d\omega \times \nonumber\\
    &\times&  \Bigg[ \frac{A}{\left( y - \frac{i\epsilon}{T} - \frac{\pi ik}{aT}\right)^{2}} - \frac{1}{4T^2}\frac{B}{\left(y - \frac{i\epsilon}{T} - \frac{\pi ik}{aT}\right)^{4}} + \frac{1}{16T^4}\frac{C}{\left(y - \frac{i\epsilon}{T} - \frac{\pi ik}{aT}\right)^{6}}\Bigg].
\end{eqnarray}
Next, we have to solve the complex integrals in $y$, for this, we choose a closed semicircular contour in the upper-half plane that surrounds the multiple poles in $y = \frac{\pi ik}{aT}$, the solutions of complex integrals present in the expression above are given by
\begin{eqnarray}\label{int complex 2}
    \int^{+\infty}_{-\infty} \frac{e^{2i\omega y}dy}{\left( y - \frac{i\epsilon}{T} - \frac{\pi ik}{aT}\right)^{2}} &=& -4\pi\omega e^{-\frac{2\pi\omega k}{aT}}, 
\end{eqnarray}
\begin{eqnarray}\label{int complex 4}
    \int^{+\infty}_{-\infty} \frac{e^{2i\omega y}dy}{\left( y - \frac{i\epsilon}{T} - \frac{\pi ik}{aT}\right)^{4}} &=& \frac{8\pi\omega^3}{3} e^{-\frac{2\pi\omega k}{aT}},
\end{eqnarray}
\begin{eqnarray}\label{int complex 6}
    \int^{+\infty}_{-\infty} \frac{e^{2i\omega y}dy}{\left( y - \frac{i\epsilon}{T} - \frac{\pi ik}{aT}\right)^{6}} &=& -\frac{8\pi\omega^5}{15} e^{-\frac{2\pi\omega k}{aT}},
\end{eqnarray}
replacing the equations (\ref{int complex 2}), (\ref{int complex 4}) and (\ref{int complex 6}) in (\ref{cp6sem_ints}), and using $\sum^{+\infty}_{k=-\infty} e^{-2\pi\omega/aT} = \left( 1 - e^{-2\pi\omega/aT}\right)$ and the expansion $e^{-\frac{2\pi\omega}{aT}} = 1 - \frac{2\pi\omega}{aT}$, we can then write
\begin{eqnarray}\label{cp6com_gaussianas}
    \mathcal{C}^\pm_{p^6} &=& \frac{aT}{4\pi^2} e^{-T^2\Omega^2} \Bigg[ A \int^{+\infty}_{-\infty} e^{-\omega^2}d\omega + \nonumber\\
    &+& \frac{B}{6T^2}\int^{+\infty}_{-\infty} \omega^2 e^{-\omega^2} d\omega + \frac{C}{120T^4}\int^{+\infty}_{-\infty} \omega^4 e^{-\omega^2} d\omega\Bigg],
\end{eqnarray}
where the solutions of these Gaussian-type integrals are given by
\begin{eqnarray}\label{gaussiana}
    \int^{+\infty}_{-\infty} e^{-\omega^2}d\omega = \sqrt{\pi},
\end{eqnarray}
\begin{eqnarray}\label{w2gaussiana}
    \int^{+\infty}_{-\infty} \omega^2 e^{-\omega^2}d\omega = \frac{\sqrt{\pi}}{2},
\end{eqnarray}
\begin{eqnarray}\label{w4gaussiana}
    \int^{+\infty}_{-\infty} \omega^4 e^{-\omega^2}d\omega = \frac{3}{8}\sqrt{\pi},
\end{eqnarray}
therefore, substituting the equations (\ref{gaussiana}), (\ref{w2gaussiana}) and (\ref{w4gaussiana}) in (\ref{cp6com_gaussianas}), and then substituting the definition expressions of $A$, $B$ and $C$, and finally simplifying the result, we obtain
\begin{eqnarray}
    \mathcal{C}^\pm_{p^6} &=& \frac{\overline{a}\sigma\sqrt{\pi}}{4\pi^2} e^{-\sigma^2} \Bigg[ 1 + \left( \overline{a}^2 + \frac{1}{2\sigma^2}\right) \frac{\overline{\gamma}^2}{6} + \left( \frac{\overline{a}^2}{3\sigma^2} + \frac{1}{20\sigma^4} - \frac{4\overline{a}^4}{5}\right)\frac{9\overline{\gamma}^4}{16}\Bigg],
\end{eqnarray}
with $\sigma = \Omega T$ and $\overline{\gamma} = \Omega\gamma$, where we use $\delta'^2 = (1+3\kappa/4)\delta^2$. Applying the condition when $\delta \to 0$ in (\ref{cp6com_gaussianas}), we get
\begin{eqnarray}
    \mathcal{C}^\pm_{p^4} &=& \frac{aT}{4\pi^2} e^{-T^2\Omega^2} \Bigg[ \left( 1 + \frac{a^2\gamma^2}{6}\right) \int^{+\infty}_{-\infty} e^{-\omega^2}d\omega + \left( \frac{\gamma^2}{6T^2}\right)\int^{+\infty}_{-\infty} \omega^2 e^{-\omega^2} d\omega \Bigg],\nonumber\\
\end{eqnarray}
and substituting the equations (\ref{gaussiana}) and (\ref{w2gaussiana}), we obtain
\begin{eqnarray}
    \mathcal{C}^\pm_{p^4} &=& \frac{\overline{a} \sigma\sqrt{\pi}}{4\pi^2} e^{-\sigma^2} \Bigg[ 1 + \left( \overline{a}^2 + \frac{1}{2\sigma^2}\right) \frac{\overline{\gamma}^2}{6}\Bigg].
\end{eqnarray}

Thus obtain the final expressions of $\mathcal{C}^\pm_{p^4}$ and $\mathcal{C}^\pm_{p^6}$.




\end{appendices}


\bibliography{sn-bibliography}% common bib file
%\input sn-article.bbl

\end{document}